# Planetary Science by the NLSI LUNAR Team:
## The Lunar Core, Ionized Atmosphere, & Nanodust Weathering

*Science of and on the Moon* undertaken by the
*NASA Lunar Science Institute* LUNAR[1] Team

*Director:* Dr. Jack Burns, University of Colorado Boulder
*Deputy Director:* Dr. Joseph Lazio, JPL

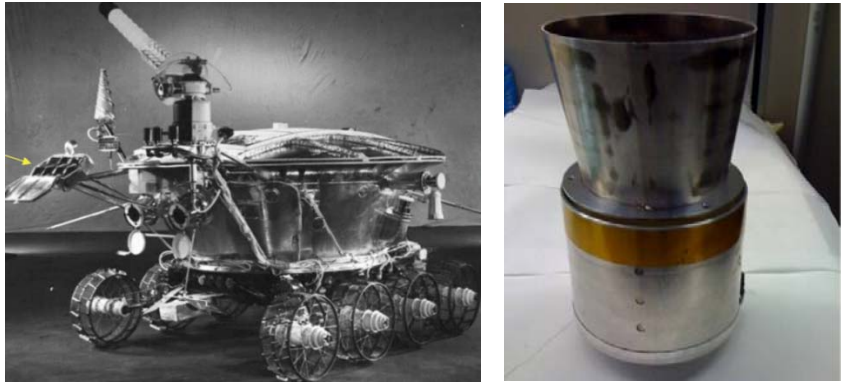

LUNAR Co-I T. Murphy (UCSD) recently performed the first lunar laser ranging (LLR) to Lunokhod 1 (left) assisted by new images from LROC. This added retroreflector is providing new libration measurements to constrain the Moon's fluid core and will complement GRAIL's measurements of the lunar interior. With the next generation corner cube retroreflector (right) being developed by LUNAR Co-I D. Currie (UMd), 100 µm LLR ranging accuracies will precisely evaluate the properties of the Moon's inner solid core and the core-mantle boundary. Such measurements could be made on a Google Lunar X- Prize flight in <2 years.

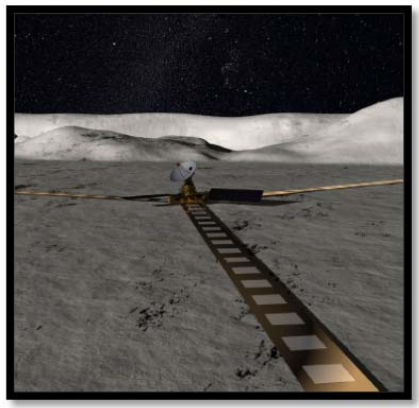 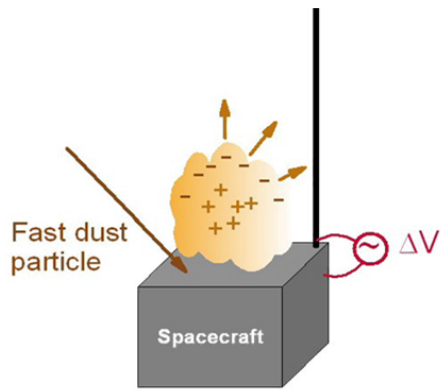

LUNAR Co-I J. Kasper (CfA) has proposed that polyimide low frequency antenna arrays (left) will be a powerful detector of nanometer-size dust grains embedded in the solar wind, recently discovered by STEREO electric field antennas (right). Such small, high kinetic energy grains may be an important new element in the weathering of airless bodies such as the Moon and asteroids.

---

[1] The Lunar University Network for Astrophysics Research (LUNAR, http://lunar.colorado.edu) is funded by the NASA Lunar Science Institute (http://lunarscience.arc.nasa.gov/) via Cooperative Agreement NNA09DB30A.

# Summary

The Lunar University Network for Astrophysics Research (LUNAR) undertakes investigations across the full spectrum of science within the mission of the NASA Lunar Science Institute (NLSI), namely science of, on, and from the Moon. The LUNAR team's work on science of and on the Moon, which is the subject of this white paper, is conducted in the broader context of ascertaining the content, origin, and evolution of the solar system.

**The interior structure and composition of the Moon**, particularly of its core, remains poorly constrained. In turn, the size and state of the core (fluid and solid components) reflect processes that occurred at the time of the formation of the Earth-Moon system, including the likely giant impact responsible for the formation of the Moon. Advancing the state of knowledge of the interior structure was recognized as an important science objective in the Planetary Sciences Decadal Survey report entitled *Visions and Voyages for Planetary Science in the Decade 2013–2023*. Lunar laser ranging provides precision measurements of the Earth-Moon distance approaching the 1 mm level, and the LUNAR team is developing the technology to advance the precision to the <100 μm level. At these levels of precision, variations in the Moon's librations are easily detectable, and, even at the current precision level using retroreflectors emplaced during the *Apollo* missions, lunar laser ranging uniquely constrains the size of the liquid core to be approximately ≈400 km in radius, with the specific value depending upon the composition of the core. Other work being undertaken tracks the influence of tides, heat dissipation, and the orbital evolution of the Earth-Moon system.

The **lunar atmosphere** is the exemplar and nearest case of a surface boundary exosphere for an airless body in the solar system. The *Visions and Voyages for Planetary Science in the Decade 2013–2023* Decadal Survey noted that understanding the evolution of exospheres, and particularly their interaction with the space environment, remains both poorly constrained and requires observations at a variety of different bodies. Determining and tracking the properties of the lunar atmosphere both robustly and over time requires a lunar-based methodology by which the atmosphere can be monitored over multiple day-night cycles from a fixed location(s). Relative ionospheric opacity measurements or *riometry*, measures the amount of power received at different radio frequencies and directly determines the density of the (ionized) atmosphere. The LUNAR team has been developing the technology for a future lunar-based radio telescope, however, the same technology is also applicable to a lunar riometer that could be deployed on a future lander, either flown by NASA or a commercial entity (e.g., a Google Lunar X-Prize competitor).

The **interplanetary medium is pervaded by dust** from a variety of sources, including small bodies, the inner planets, and interstellar space that may be a **key element in the weathering of airless bodies** in the solar system. Recent work on interplanetary dust by members of the LUNAR team has revealed a substantial population of nanometer-size dust, or *nanodust*, with fluxes hundreds of thousands of times higher than the better understood micron-sized dust grains. This nanodust tends to move with the speed of the solar wind (i.e., 100's of km/sec) as opposed to more typical Keplerian speeds (i.e., 10's of km/sec). Since impact damage grows faster than the square of the impact speed for high speed dust, nanodust could be an important and previously unrecognized contributor to space weathering. The same technology for a future lunar-based radio telescope would also be a superb tool for measuring the distribution of dust particles as a function of size in interplanetary space, and ultimately for understanding how dust modifies the surfaces of planets and other airless objects in the solar system.

# 1. Interior Structure and Composition of the Moon via Lunar Laser Ranging

*Project Leaders:*   Dr. Douglas Currie, University of Maryland, College Park
Dr. Stephen Merkowitz, Goddard Space Flight Center
Dr. Thomas Murphy, University of California, San Diego

### a. *Recommendations from the Planetary Sciences Decadal Report Vision and Voyages for Planetary Science in the Decade 2013-2022:* **Lunar Geophysical Network**

The Planetary Sciences Decadal Survey emphasized the unique role that the Moon's interior plays in understanding the formation and evolution of the Earth-Moon system. The Survey recommended that "Deploying a global, long-lived network of geophysical instruments on the surface of the Moon to understand the nature and evolution of the lunar interior from the crust to the core (will) allow the examination of planetary differentiation that was essentially frozen in time at some 3-to-3.5 billion years ago. *Such data (e.g., seismic, heat flow, laser ranging, and magnetic field/electromagnetic sounding) are critical to determining the initial composition of the Moon and the Earth-Moon system.*"

The Survey went on to state that important scientific objectives include: "Determine the size of structural components (e.g., crust, mantle, and core) making up the interior of the Moon, including their composition and compositional variations to estimate bulk lunar composition and how they relate to that of Earth and other terrestrial planets, how the Earth-Moon system was formed, and how planetary compositions are related to nebular condensation and accretion processes."

Lunar Laser Ranging (LLR) plays a unique and critical role in determining the nature of the Moon's core. The research by the LUNAR team is advancing our knowledge of the lunar interior using current (Apollo and Soviet era) retroreflectors on the Moon's surface and is actively advancing the technology for a new generation of retroreflectors that could be flown in a few years by Google Lunar X-Prize teams or on new NASA missions.

### b. *Summary of Lunar Laser Ranging Activities by the LUNAR Team*

The Lunar Laser Ranging component of the LUNAR team has addressed the exploration of the interior of the Moon via the preparation of a retroreflector package (i.e., the Lunar Laser Ranging Retroreflector Array for the 21$^{st}$ Century or LLRRA-21) that can be ready for launch within 18 months (after the receipt of funds). This will provide several orders of magnitude improvement in the science already supplied by the LLR program that is using the existing retroreflectors from the Apollo era. The latter program has provided extensive information on the liquid core (existence, size and shape) data on the Core-Mantle Boundary and other aspects of the lunar interior.

The LLRRA-21 will complement the GRAIL mission. While the GRAIL investigation is strongest near the lunar surface, the LLR program using the LLRRA-21 is strongest for the deep interior of the Moon, for example looking for the inner solid core. By taking advantage of the Google Lunar X-Prize, this complementary mission can be achieved at a cost of a few percent of the cost of the GRAIL mission.

### c. Background on Lunar Laser Ranging (LLR) Technique

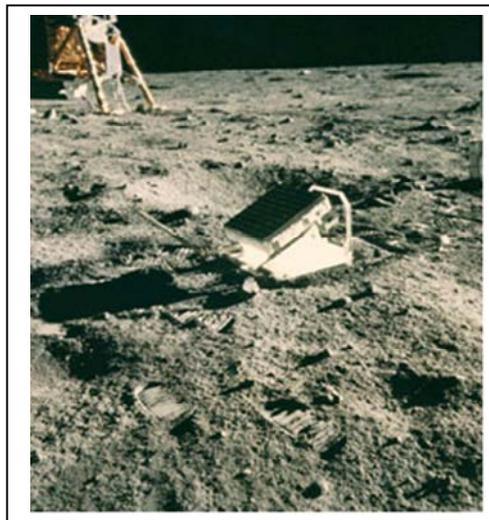

The University of Maryland, one of the lead member institutions of LUNAR, was Principal Investigator on the development of the retroreflector arrays that were carried to the Moon during the Apollo missions (Bender, Currie, et al. 1973). These provided fixed point targets on the lunar surface that sent the laser signal directly back to the telescope which was being used as the laser transmitter as well as the return signal receiver. They were designed to operate in both the lunar night and the lunar day with a long lifetime. They are still producing excellent data for the analysis of the interior of the Moon.

Most of the ranging to the Apollo arrays has been done by a limited number of stations, in particular, McDonald (107-in), McDonald (MLRS), the French, Maui, and now APOLLO (Apache Point Observatory Lunar Laser-ranging Operation). These stations have collected data mostly from the Apollo 15 retroreflector, since it has a stronger return. Often when stations go down, the coverage is rather sparse. However, even with this, the existence of a long data series has led to the ability to isolate various frequencies with long periods in the libration history of the Moon, which are produced by a variety of interior phenomena and, thus, permits the extraction of a large body of science.

During the interval from March 16, 1970 to November 22, 2008, a total of 16,941 ranges were made. Ranges were processed from McDonald Observatory, Texas (6,523 ranges), Observatoire de la Côte d'Azur (OCA), France (9,177), Haleakala Observatory, Hawaii (694), Apache Point Observatory, New Mexico (536), and Matera, Italy (11).

LLR data are archived and available to the public by the International Laser Ranging Service (ILRS) (Pearlman et al., 2002). The most accurate data and science analysis have been done by the Jet Propulsion Laboratory (JPL). As the accuracy has been improved to the 20 mm level, there has been continual improvement in the software and continual improvement in the extracted science. Various papers have addressed the current and future science output of the Lunar Laser Ranging Program (Williams et al. 2010, Ratcliff et al. 2008).

The Apache Point Observatory (APOLLO) is a key high accuracy addition to the LLR network (Murphy et al., 2009, Battat et al. 2009). The APOLLO data has potential accuracy approaching 1 mm by collecting thousands of returns obtained with a large astronomical telescope.

### d. Lunar Interior Science Results of the Lunar Laser Ranging Program (LLRP)

Our core data set for the evaluation of the lunar interior is the historical record of the librations. This is most effectively obtained by regularly observing all five retroreflectors in a short period of time of a few hours (which APOLLO is currently able to do). This cancels a number of program and observational systematic errors. The fact that ranging to either Apollo 11 or 14 is more difficult for most stations has resulted in a reduced accuracy in the libration historical record and thus knowledge of the lunar interior. With the recent successful lunar ranging to the Lunokhod 1 by LUNAR Co-I T. Murphy, libration measurements are expected to dramatically improve and, thus, provide important new constraints on the lunar core.

We now address some of the leading science that has been extracted from the forty year history of the librations (see e.g., Williams et. al. 2008).

#### i. Liquid Core: Fluid Core Moment of Inertia

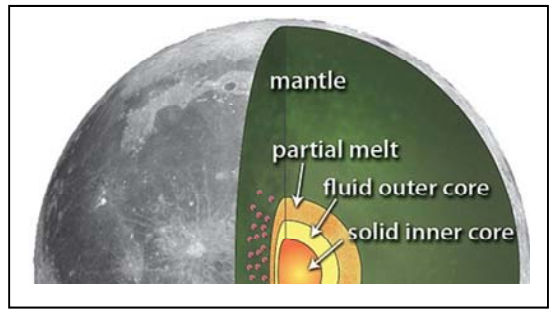

The fluid core moment of inertia is the latest lunar geophysical parameter to emerge from the LLR analysis. Sensitivity comes because the orientations of both mantle and fluid core follow the slow motion of the ecliptic plane, while the core has diminished response to faster variations. The solution for the ratio of fluid moment to total moment gives $C_f/C = (12\pm4) \times 10^{-4}$. For a uniform liquid iron core without an inner core this value would correspond to a radius of 390±30 km, while for the Fe-FeS eutectic the radius would be 415 km. Those two cases would correspond to fluid cores with 2.4% and 2.2% of the mass, respectively.

#### ii. Tides

There is information from the Apollo seismometers on the elastic properties of the lunar crust, but that information does not extend to the lower mantle and core. Of the three Love numbers, $l_2$ is least sensitive to the deep zones so with LLR we can solve for $k_2$ and $h_2$ while fixing $l_2$ at a model value of 0.0105. Solutions give $k_2 = 0.0199\pm0.0025$ and $h_2 = 0.042\pm0.008$.

#### iii. Dissipation from Tides and the Core

The key to separating the two causes of gravitational dissipation was the detection of small physical libration effects of a few milliarcseconds (mas) size. The tidal time delay and the core-mantle boundary (CMB) dissipation are both effective at introducing a phase shift in the precessing pole direction. The solution gives dissipation from the CMB and tides. Both are strong contributors to the 0.27" offset of the precessing rotation pole from the dissipation-free pole, equivalent to a 10" shift in the node of the equator on the ecliptic plane (Williams et al., 2001).

### iv. Core Oblateness

The detection of the oblateness of the fluid-core/solid-mantle boundary is independent evidence for the existence of a liquid core. In the first approximation, CMB oblateness influences the tilt of the lunar equator to the ecliptic plane. Torque from an oblate CMB shape depends on the product of the fluid core moment of inertia and the CMB flattening, $fC_f=(C_f-A_f)$. The LLR solution gives $f=(C_f-A_f)/C_f=(2.0\pm2.3)\times10^{-4}$. For a 390 km core radius the flattening value would correspond to a difference between equatorial and polar radii of about 80 m with a comparable uncertainty (Williams et al. 2008, 2009).

### v. Free Librations

Gravitational dissipation has been determined by LLR from both tidal flexing and the fluid/solid interaction at the core/mantle boundary. Dissipation introduces a phase shift in each periodic component of the forced physical librations and substantial motions are found for two of the modes. The longitude mode is a pendulum-like oscillation of the rotation about the (polar) principal axis associated with moment C. The period for this normal mode is 1056 d =2.89 yr. The lunar wobble mode is analogous to the Earth's polar motion Chandler wobble, but the period is much longer and the path is elliptical with a 74.6 yr period and the amplitudes are 3.3"x8.2" (28 m x 69 m). The computed damping time is about 106 yr. The mantle free precession of the equator (or pole) has an 81 yr period. An amplitude of 0.03" is found for this mode. The expected damping time is $2\times10^5$ yr. The fluid core free precession of the fluid spin vector has an expected period >170 yr, as previously discussed under Core Oblateness; it would be 197 yr (Rambaux et al. 2008).

### vi. Search for a Solid Inner Core

It is reasonable to expect that the Moon would have a solid core interior to the fluid core, but it remains undetected with high confidence. The phase diagram for Fe-FeS shows that cooling of fluid alloys of iron and sulfur would freeze out part of the iron while concentrating sulfur compounds in the fluid. There is no direct evidence for a solid inner core. In principle, an inner core might be detected through its influence on physical librations or gravity, or through seismology. Any detection would establish the last major unit of the Moon's structure.

To look for inner core effects, the post-fit LLR residuals for each retroreflector array have been analyzed to produce spectra. The Apollo 11 and 14 arrays are near the equator, so they will be most sensitive to longitude librations. The Apollo 15 array, well north of the equator with a small longitude, provides the most sensitivity to latitude librations. The Lunokhod 2 array is sensitive to both longitude and latitude librations, but the smallest number of observations (477 is 2.8%) gives this array the noisiest spectra. The detection and measurement of the Moon's inner core will be a major accomplishment for any technique. For LLR, it is a future possibility.

### vii. Orbit Evolution

Dissipation of gravitational energy in the Moon and Earth causes slow changes in the lunar orbit. The semimajor axis and eccentricity increase with time and the inclination decreases. Dissipation in the Moon also deposits heat in the Moon. This is a minor effect now, but could have been much more important when the Moon was closer to the Earth. Here we summarize the orbit changes.

The orbit evolution is addressed by a determination of the apparent acceleration $(dn,/dt)$, in the lunar mean motion n. An independent LLR analysis for total dn/dt of $-25.858"/century^2$ which gives very good agreement with the JPL determination mean longitude acceleration, using the DE421 ephemeris, of $-25.85"/century^2$ given here. The DE421 value corresponds to a 38.14 mm/yr rate for the changes in the semimajor axis. Accounting for the difference in de/dt from the simple LLR integration model and the more complete Earth model, the unexplained eccentricity rate is $(0.9\pm0.3) \times 10^{-11}$ /yr, equivalent to an extra 3.5 mm/yr in the perigee rate.

### e. Technical Implementation of next generation Lunar Laser Ranging Program

#### i. Next Generation Lunar Retroreflectors

##### 1. Solid Cube Corner Reflectors (CCRs)

The Apollo Retroreflectors are arrays of Cube Corner Reflectors (CCRs) which consist of solid prisms composed of fused silica. They have performed well over the past four decades, still providing highly valuable data. In addition, many hundreds of solid CCRs are in orbit on satellites and again, these have performed well over the decades. Thus, the use of solid CCRs has a long and excellent heritage. The LLRRA-21 proposed by the University of Maryland and developed within the LUNAR program, consists of a single CCR that has a diameter that is 2.6 times larger than the Apollo CCRs (Currie et al. 2011a, Dell'Agnello et al. 2011). The use of a single CCR removes the pulse spreading due to the lunar librations that plagues the accuracy that may be obtained with for a single return from the Apollo arrays. The magnitude of the signal return on Earth depends upon the uniformity of the index of refraction and therefore on the temperature gradients within the CCR. Thus, the larger CCR has required a careful thermal analysis and optical/thermal/vacuum testing to validate the thermal design

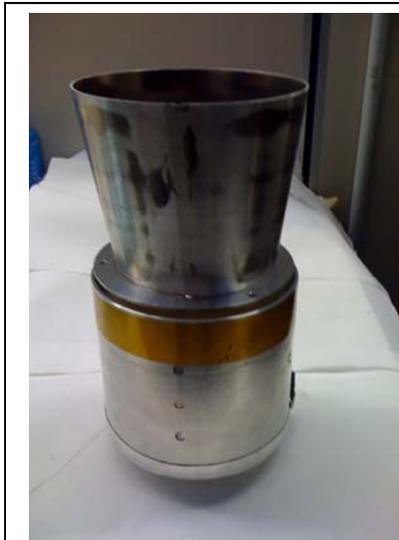

Package that has been testing the thermal/vacuum system (SCF) at Frascati, Italy.

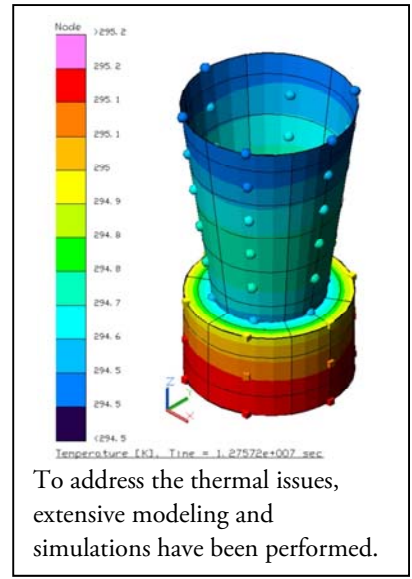

To address the thermal issues, extensive modeling and simulations have been performed.

and the simulation programs. This has been accomplished and has been used to select the various configuration issues and the thermal coatings. With the addition of thermal vacuum testing, the LLRRA-21 is currently at a Technology Readiness Level (TRL) of 6.5 and is expected to be at 7.0 after tests later this year.

The ultimate accuracy of the lunar emplacement ranges from 2 mm to 0.1 mm, depending upon the method of deployment. The various deployment approaches are currently being investigated, in analysis and in hardware implementation. (Currie 2011b; Zacny & Currie 2011a; Currie 2011c). The return signal level from the Apollo arrays has decreased and is now at a level of 10-100 below the return expected when initially deployed, depending upon the phase in a lunation. The causes of this are being addressed to assure that such a problem does not occur with the LLRRA-21 (Currie, Horanyi, et.al 2011). Finally, the LLRRA-21 package has been extensively tested in the INFN thermal Vacuum chamber (Dell'Agnello et al. 2011; Currie, Dell'Agnello, & Delle Monache 2011; Dell'Agnello, Delle Monache, & Currie 2011). A Google Lunar X-Prize flight would offer the opportunity for lunar emplacement within the next 2 years.

## 2. Hollow Cube Corner Reflectors

Within the LUNAR program, a parallel effort at the Goddard Space Flight Center by LUNAR Co-I. S. Merkowitz is addressing the use of open or hollow retroreflectors for use as the next generation lunar laser retroreflector. The open retroreflector has a number of theoretical advantages and there is an active research program at GSFC to demonstrate the feasibility of this approach and evaluate the theoretical advantages of the open retroreflector. The objective is to demonstrate that this technology can meet the very critical environmental and stability requirements for long term operation on the surface of the Moon.

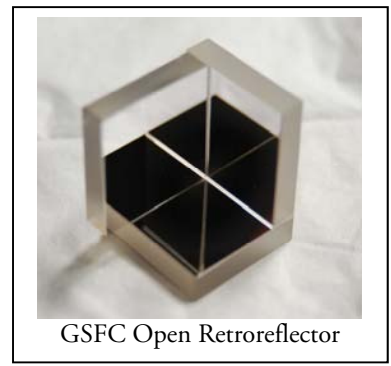

GSFC Open Retroreflector

## ii. Return Signal Level and Ground Stations

The LLRRA-21 provides sufficient signal so that many stations that are dedicated to laser ranging could participate in the LLR program and, thus, obtaining many observations per month at the millimeter level. Although the APOLLO (Apache Point Observatory Lunar Laser-ranging Operation) currently obtains millimeter level observations, due to other uses of the telescope, this occurs only a few times a month. Due to the ability of satellite ranging stations to obtain millimeter accuracy with a few returns and due to the fact that the satellite stations conventionally range many times a month, the LLRRA-21 could result in two orders of magnitude more ranges per month.

## iii. Software

The current analysis software is appropriate at the 20 mm level, but will require much work to reach the 1 mm level. To address this, under LUNAR auspices an international meeting was held recently in Cambridge, MA. French, German, and US analysis groups were represented and presented data on the orientation of the Moon as derived from the same data set. This indicated that the JPL program is at present the best selection. The other conclusion was that the APOLLO data was shown to be at least as good as two millimeters in accuracy.

### f. Science Objectives for the Next Generation LLRP

*The new reflectors developed under the auspices of the LUNAR team will allow two orders of magnitude improvement in the single shot ranging accuracy*, which in turn will allow a much higher data rate and a high data rate for observation of multiple reflectors in a short interval. The increased accuracy, the greater number of observations per month, and the increase in the number of observing stations will allow a great improvement over the accuracies described above, which already were the best probe of various parameters of the interior of the Moon.

The issue of predicting the expected accuracies is one of the topics being pursued by a collaboration between the University of Maryland, the National Astronomical Observatory of Japan (NAOJ) and INFN of Frascati, Italy. H. Noda of the NAOJ is working with James Williams of JPL to perform these projections. This is a difficult project due to the long computer runs, due to the fact that the software must be upgraded in order to address what will happen with 1 mm data, and due to the limited access available to H. Noda (Williams, Boggs, & Noda 2010).

### g. Technology Objectives for the Next Generation LLRP

#### a. Drilling in the Regolith

Many of the science objectives for future NASA missions require drilling into the regolith, for example heat flow, core sampling as well as the optimal emplacement of the Lunar Laser Ranging Retroreflector for the 21$^{st}$ Century (LLRRA-21). New drilling technologies have emerged since the Apollo missions but have not had the opportunity of being tested "in situ" on the real regolith. Under the auspices of LUNAR, University of Maryland has been working with Astrobotics and Honeybee in order to define a LLRRA-21 mission on the Google Lunar X-Prize (GLXP) program. This mission would consist of using the pneumatic drill developed by Honeybee (funded by LUNAR) on the lander being developed by Astrobotics (and/or the landers for the Next Giant Leap and/or other GLXP Teams). Thus, this whole new concept of pneumatic drilling may be tested in the real environment within the next few years. This program (developed within the LUNAR program funded by NLSI) will provide information that will be of great value for missions like Lunar Geophysical Network as well as other missions to land on the surface of airless bodies.

#### b. Selenodesy

An accurate grid or selenographic coordinate system will be critical in future robotic and/or manned missions. Lunokhod 1 had never been ranged to since the coordinates of the final resting place were not sufficiently accurate. Using the high resolution imagery, LRO identified the location of Lunokhod 1 and these coordinates were given to LUNAR Co-I T. Murphy for use at the APOLLO

Station. This allowed laser returns to be obtained from Lunokhod 1 for the first time. However, the LLR position of Lunokhod 1 was different from the LRO coordinates by 100 meters. Thus, this new data point should allow a very significant upgrade to the seleodetic coordinate system being used by LRO. Additional retroreflectors will serve to tie down the coordinate system in a variety of new locations.

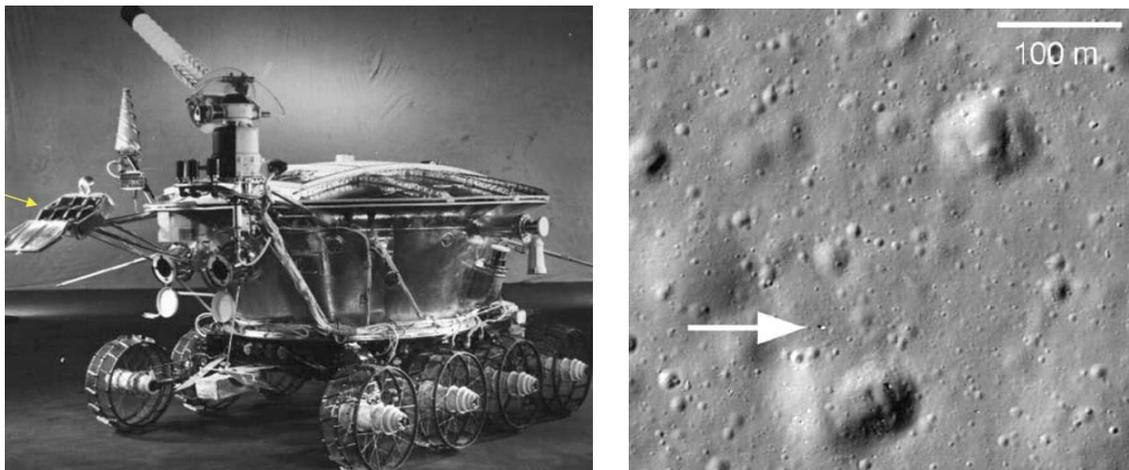

Lunokhod 1 Lander (left). LRO LROC image of the Lunokhod 1 lander site (right). The APOLLO lunar laser ranging operation recently detected strong laser return signals from Lunokhod 1 but found a lateral offset of 100 meters in the location of the Lunokhod. Next generation retroreflectors will significantly improve the selenographic coordinate system in preparation for human surface operations.

### h. Relation to Other NASA Lunar Missions

#### i. Apollo

The current arrays were placed during the Apollo 11, 14 and 15 missions. They are still operational and providing information on the lunar interior. This has been especially productive due to the long data series, far longer (40 years) than any other Apollo experiment. This has allowed the determination of lunar interior properties that are characterized by the long period (18.6 years) in the lunar librations. However, the single shot accuracy on the Apollo arrays is limited by the use of multiple CCRs in an array that rotates with the lunar librations. This leads to an uncertainty of the order of 20 mm.

The next generation retroreflectors (e.g., LLRRA-21) will give a signal return that is larger than the return obtained with Apollo 15 so all of the currently operating lunar laser ranging stations can expect useful signal strength. In addition, the new retroreflectors will have sufficient signals for other stations with sub-meter aperture telescopes to join the group doing the ranging, and provide more than a few observations per month at the mm accuracy level, which is the current situation.

#### ii. LRO

One of the scientific objectives of the Lunar Reconnaissance Orbiter (LRO) is to make details maps of the composition and structure of the lunar surface, using a variety of complementary methods such as altimetry, imaging, UV, IR and neutron observations. Some of these instruments have a ground resolution of centimeters. For this information to be valuable to other observers and other experiments, one must locate the observations within a selenodetic coordinate system. This also applies to future robotic and manned landing missions that need to know where they are located with respect to LRO data.

LRO located the Lunokhod 1 rover with a retroreflector array. These coordinates were given to LUNAR Co-I T. Murphy to use for lunar laser ranging from the APOLLO station. Conducted within the LUNAR program, these observations were successful and will provide data to improve offset errors in the selenodetic coordinate system from 100 meters to less than a meter. *Since Lunokhod 1 is the furthest retroreflector from the sub-earth point, it will be extremely valuable in determining the librations of the Moon, which will translate into improved information on the above areas and of the science of the interior of the Moon.*

### iii. GRAIL

The GRAIL mission operates by sampling the horizontal gradient of the lunar gravity field at an altitude of 50 to 100 km. This gives an excellent determination of the gravitational properties of the "mass-cons" (i.e., mass concentrations) and the gravitational or mass properties of the upper crust and upper mantle. However, in the lower mantle, at the core-mantle boundary, the liquid core and the conjectured inner solid core, the signal-to-noise ratio to evaluate the science properties drops considerably.

As illustrated by the fact that the current determination of the existence, size and shape of the liquid lunar core has been determined from the LLR program, *LLR is possibly the best tool for evaluation of the physical properties of the deep lunar interior.* There are two reasons for the high sensitivity of the LLR to the deep interior. The first is the very high accuracy of the evaluation of the rotational parameters of the Moon. These rotational parameters or libration parameters are affected by the "sloshing" and the partially independent rotation of both the liquid core and the conjectured inner solid core. The second is that many of these parameters have periods of 6 or more years. Thus, *the demonstrated long lifetime of the LLR program provides a strong complement to the GRAIL mission* with its sampling of parameters over a lifetime of two years. Further, the Google X-Prize emplacements would allow the lunar program to start simultaneously with the GRAIL mission. Currently, a collaboration is starting with the GRAIL personnel in order to optimize the science obtained by combining the observations of both missions.

## 2. The Lunar Ionosphere/Ionized Atmosphere

*Project Leader:*     Dr. Joseph Lazio, JPL

The lunar atmosphere is the exemplar and nearest case of a surface boundary exosphere for an airless body in the solar system, and the *Visions and Voyages for Planetary Sciences in the Decade 2013–2023*

Decadal Survey noted the importance of tracking the evolution of exospheres, particularly in response to the space environment. Determining and tracking the properties of the lunar atmosphere both robustly and over time requires a lunar-based methodology by which the atmosphere can be monitored over multiple day-night cycles from a fixed location(s), such as a *lunar relative ionosphere opacity meter* (*riometer*).

### a. Science Background

Speculations on and attempts to measure a lunar atmosphere extend back at least to the 18$^{th}$ Century (e.g., Schroeter 1792; Challis 1863). Prior to the Space Age, radio astronomical measurements of lunar occultations (Elsmore 1957) led to the suggestion of a tenuous lunar atmosphere, which quickly led to various potential sources being identified (Edwards & Borst 1958), including one still considered plausible, namely the outgassing of primordial gasses. It was also recognized that the tenuous nature of the lunar atmosphere meant that it would be strongly influenced by its environment, e.g., by its exposure to the solar wind (Herring & Licht 1958). By the time of the Apollo missions, the major properties of the lunar atmosphere were being elucidated as well as the recognition that even modest human activity could produce substantial alterations (Milford & Pomilla 1967).

Reviews of the state of knowledge of the lunar atmosphere at the close of the Apollo era indicate significant advances in knowledge of the composition, sources and sinks, and influences on the lunar atmosphere but also significant questions about all of these topics (e.g., Johnson 1971; Hodges et al. 1974). Exposed to both the solar and interstellar radiation fields, the daytime lunar atmosphere is mostly ionized, and we shall use the terms atmosphere and ionosphere interchangeably. Enduring questions include the density and vertical extent of the ionosphere and its behavior over time, including modification by robotic or crewed landers.

ALSEP measurements during the Apollo missions found a photoelectron layer near the surface with electron densities up to $10^4$ cm$^{-3}$ (Reasoner & O'Brien 1972). Further, dual-frequency radio occultation measurements from the Soviet Luna spacecraft suggest that the ionosphere's density is both highly variable and can extend to significant altitudes, exceeding $10^3$ cm$^{-3}$ well above 10 km (Figure 1). However, the interpretation of the Luna data is model dependent, as Bauer (1996) concluded that the Luna data were consistent with no significant lunar ionosphere.

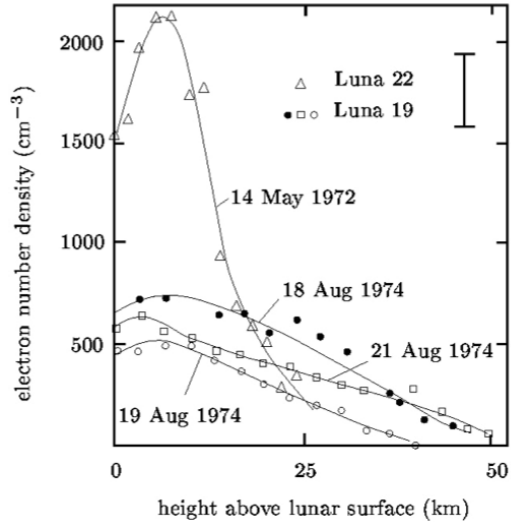

Figure 1. Lunar ionosphere electron densities derived from dual-frequency radio occultation measurements during the Luna 19 and Luna 22 missions (Vyshlov 1976; Vyshlov & Savich 1978).

In addition to an ion or molecular component to the plasma layer above the Moon's surface, there are reports of a "horizon glow" from both crewed and robotic missions (e.g., Rennilson & Criswell 1974; Zook & McCoy 1991). There is widespread agreement that this "horizon glow" is likely due to electrostatically charged dust that is levitated above the surface. Such a component would also contribute free electrons to the atmosphere.

More recently, there have been a series of spacecraft-based remote sensing efforts to measure the lunar ionosphere. Pluchino et al. (2008) performed dual-frequency (2200 and 8400 MHz) lunar occultation observations of the SMART-1, *Cassini*, and Venus Express spacecraft. One of the experiments on the Japanese SELENE (KAGUYA) mission used a series of dual-frequency measurements (at 2200 and 8500 MHz) in an effort to detect the lunar ionosphere (Imamura et al. 2008). In general, an increase in the electron density on the solar illuminated side of the Moon has been observed, consistent with the expectations for the presence of a lunar ionosphere.

A significant, and acknowledged, source of systematic error for these (and any) spacecraft-based measurements is the transmission through the terrestrial ionosphere. The standard observational methodology consists of transmitting signals from the spacecraft to a (terrestrial) ground station, with the signals passing not only through the lunar ionosphere but also through the terrestrial ionosphere. With typical electron densities of $10^6$ cm$^{-3}$, the magnitude of the terrestrial signal is much larger than the magnitude of any potential lunar signal. Spacecraft-based remote sensing ultimately is, and will be, limited by the extent to which the (much larger) terrestrial ionospheric contribution can be estimated and removed.

Further, it is recognized that the density of the lunar ionosphere should vary dramatically, due to a variety of influences including the diurnal solar illumination changes, immersion in the Earth's magnetosphere, changes over a solar cycle, and even exhaust from spacecraft and landers. However, many of these spacecraft measurements of the lunar ionosphere suffer from relatively short mission durations, in some cases consisting of only a single epoch.

## b. Technical Approach

The principle underlying *relative ionospheric opacity measurements* (riometry) is that the refractive index of a fully or partially ionized medium (plasma) is a function of frequency and becomes negative below a characteristic frequency known as the *plasma frequency*. At frequencies below the plasma frequency, an electromagnetic wave cannot propagate through the medium and is reflected upon incidence. The plasma frequency is given by

$\omega_p = (4\pi n_e e^2/m_e)^{1/2}$

where $n_e$ is the electron density, e is the charge on the electron, and $m_e$ is the mass of the electron. With $\omega_p = 2\pi f_p$, and substituting for physical constants,

$f_p = 9$ kHz $(n_e/1$ cm$^{-3})^{1/2}$.

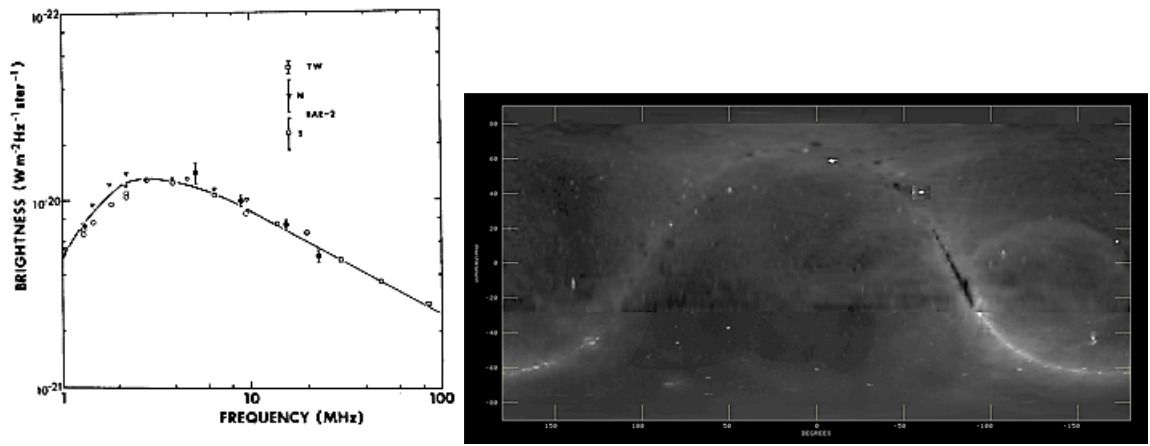

**Figure 2.** (Left) Spectrum of the Milky Way Galaxy's synchrotron emission between 1 and 10 MHz (Cane 1979). This emission is both well characterized and sufficiently broadband to be used for either terrestrial or lunar riometry. (Right) An all-sky model at 10 MHz, constructed by synthesizing a variety of low-frequency sky measurements (Polisensky 2007). The prominent band through the image is the plane of the Milky Way Galaxy.

A riometer exploits this characteristic of a plasma to measure the ionospheric density. If a broadband reference emitter with a known spectrum is observed through the ionosphere, the ionosphere's plasma frequency, and in turn the (peak) ionospheric density, can be determined from the frequency at which absorption occurs. By monitoring the plasma frequency, a riometer can track changes in the ionospheric density over time. In practice, the most commonly used reference emitter is the non-thermal radio emission from the Milky Way Galaxy, which results from synchrotron emission generated by relativistic electrons spiraling in the Galaxy's magnetic field. This Galactic emission has the favorable properties of being both extremely well characterized and constant in time (Figure 2).

A *relative ionospheric opacity meter* (riometer) is conceptually simple, consisting of an antenna, a receiver, and a data storage unit. Riometers have been used for decades, in remote and hostile locations, for tracking the properties of the Earth's ionosphere. Riometers could be deployed robotically, by an astronaut, or by an astronaut telerobotically operating a rover. Riometers could form a science enhancement option for the geophysical focus of the Lunar Geophysical Network described in the NRC Decadal Survey *Vision and Voyages for Planetary Science in the Decade 2013–2022*. In addition to basic lunar science, a riometer also supports lunar exploration by tracking the modification of the lunar atmosphere by exhaust from landers.

# 3. Weathering the Moon's Surface via Interplanetary Nanodust Impacts: Measurements with a Radio Array

*Project Leader:* Dr. Justin Kasper, Harvard-Smithsonian Center for Astrophysics

Interplanetary space is pervaded by dust with sizes ranging from nanometers to tens of microns and larger. Recent work on interplanetary dust has revealed a substantial population of nanometer-size dust, or nanodust, with fluxes hundreds of thousands of times higher than better understood micron-sized dust grains. This nanodust tends to move with the speed of the solar wind, or at hundreds of kilometers per second, as opposed to more typical Keplerian speeds of tens of kilometers per second. Since impact damage grows faster than the square of the impact speed for high speed dust, this nanodust can generate significant damage when it impacts an object such as a spacecraft, or a planet, or the Moon, as shown in Figure 3. In this section we describe how a low frequency radio array is ideal for measuring the distribution of dust particles as a function of size in interplanetary space, and ultimately for understanding how dust modifies the surfaces of planets and other objects in the solar system.

Dust has many sources, including collisions between asteroids, escaping gas from comets, and condensation within the solar atmosphere. Additional dust streams into the solar system from interstellar space. The size, speed, and mass distribution of dust in interplanetary space and its variation with time tell us about the history of these sources. Measurements of dust properties have been performed with dedicated dust instruments specifically designed to characterize dust particles (Grun et al., 1992; Srama et al., 2004). More recently, it has been shown that space-based radio receivers can also be used to measure dust. These radio instruments function by measuring the electrical signals produced when dust grains impact objects at high speed and create expanding clouds of plasma (e.g. Meyer-Vernet, 2001, Zaslavsky et al. 2011, and references therein). Work on the use of radio receivers for studying dust, including recent results supported by LUNAR, have shown that radio observations have two particular strengths when it comes to conducting a survey of the interplanetary dust population. First, radio arrays are very sensitive to nanodust, a major fraction of the dust population in the solar system that produces weak signals in standard instruments. Second, a radio array is also ideal for searching for the highest mass, but rarest dust particles. This is because the entire surface area of the array, which for lunar concepts exceeds thousands of square meters, becomes a single sensitive dust detector.

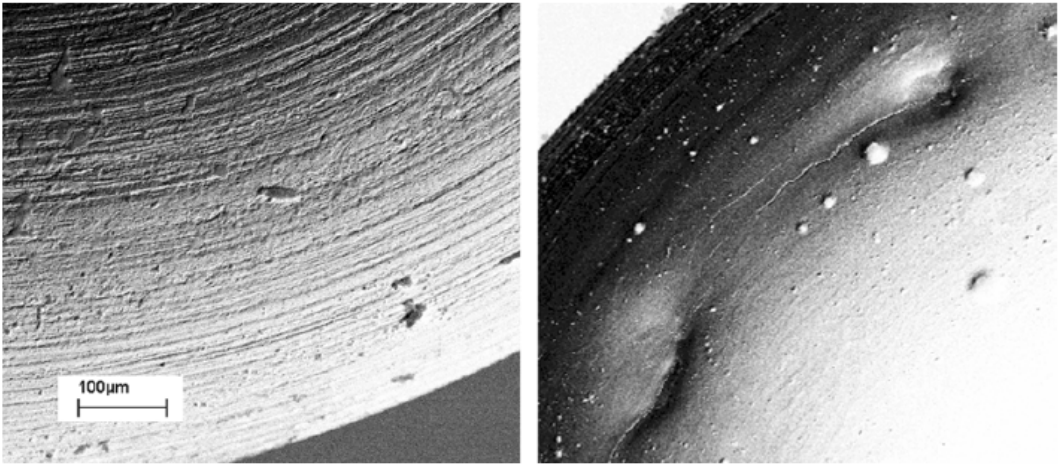

**Figure 3**: Electron microscope images of a target before (left) and after (right) being exposed to high speed dust impacts. If 1 μm dust impacts a target at 20 km/s it produces a 50 micron diameter crater. 10 nm dust is much smaller, but since it moves at 300 km/s, it can create a 1 micron crater, and is much more common than larger dust grains.

Recently, Meyer-Vernet et al. (2009) proposed that small signals seen by the electric field antennas on one the STEREO spacecraft may be due to nanometer scale dust particles. In order to determine the range of masses, and the uniqueness of radio measurements of dust properties, the LUNAR Heliophysics team led by Co-I J. Kasper analyzed dust impacts recorded by the STEREO/WAVES radio instrument onboard the two STEREO spacecraft near 1 AU during the period 2007-2010. The objective of this work was to develop an analytic set of equations to describe the evolution of a plasma plume created when dust strikes an object such as an electric field antenna, and to use these resulting equations to derive the mass distribution of the dust observed by STEREO. These results are in press (Zaslavsky et al., 2011), and were presented at 2010 American Geophysical Union and the 2011 Lunar Science Forum.

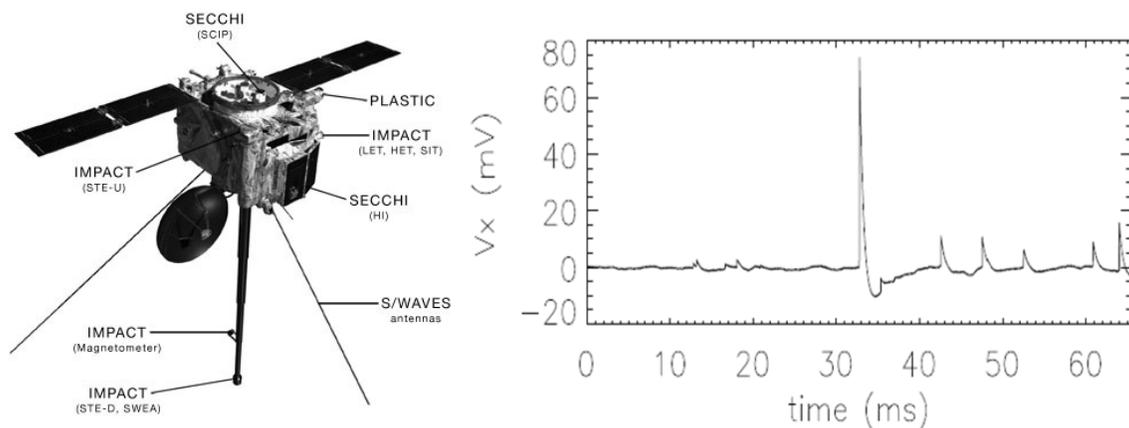

**Figure 4:** The LUNAR Heliophysics team has made use of observations by the electric field antennas on the twin STEREO spacecraft (one of which is shown on the left) to examine the radio signatures of dust impacts. Sudden changes to the spacecraft potential as measured by individual antennas (right) are produced when an interplanetary dust particle impacts the spacecraft. Since the antennas are detecting the large scale plasma plume produced by a dust impact, they are sensitive to impacts over many square meters or surface area, and therefore are capable of conducting high sensitivity searches for rare impacts.

The impact of a dust particle on a spacecraft produces a plasma cloud whose associated electric field can be detected by on-board electric antennas, as shown in Figure 4.  In our study, we used the electric potential time series recorded by the waveform sampler of the instrument. The high time resolution and long sampling times of this measurement enabled us to deduce considerably more information than in previous studies based on the dynamic power spectra provided by the same instrument or by radio instruments onboard other spacecraft. The large detection area compared to conventional dust detectors provides flux data with better statistics.  Both of these improvements on previous studies would continue with the use of a larger lunar array.

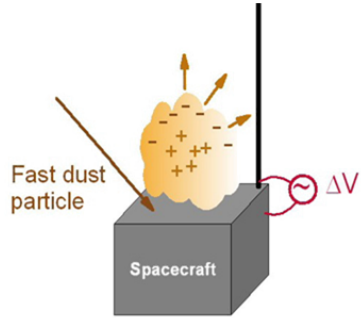

**Figure 5:** Creation of an expanding cloud of plasma surrounding a spacecraft after the impact of a dust grain.  Since ions and electrons in the cloud have the same thermal energy, the electrons expand much more quickly.

The process responsible for the signal seen by a radio antenna near the impact of a dust grain is shown in Figure 5.  When a grain impacts an object at extremely high speeds (a hyperkinetic impact), it generates a cloud of high temperature plasma of total charge approximately proportional to the mass of the grain and the speed of the grain to the 3.5 power.  Since ions and electrons in the cloud have the same thermal energy, the electrons expand much more quickly, and create a large potential drop along the antenna.  Our analysis suggests that this technique works very well for measurements that cover the mass intervals $10^{-22}$ – $10^{-20}$ kg and $10^{-17}$ – $5\times10^{-16}$ kg. The flux of the larger dust agrees with measurements of other instruments on different spacecraft, and the flux of the smaller dust grains agrees with theoretical predictions.  Figure 6 shows the variation of the dust flux in a given dust range as a function of time.  Figure 7 shows a summary of the typical dust fluxes as a function of mass.

For a lunar radio array with 3 arms of 500 m length each and average antenna width on the arms of 1 m, the surface area would be 1500 m². Given the flux distribution measured in Figure 7, this would correspond to approximately $10^3$ dust impacts per second for nanodust, and detections of the heavy 10 micron dust several times a minute.

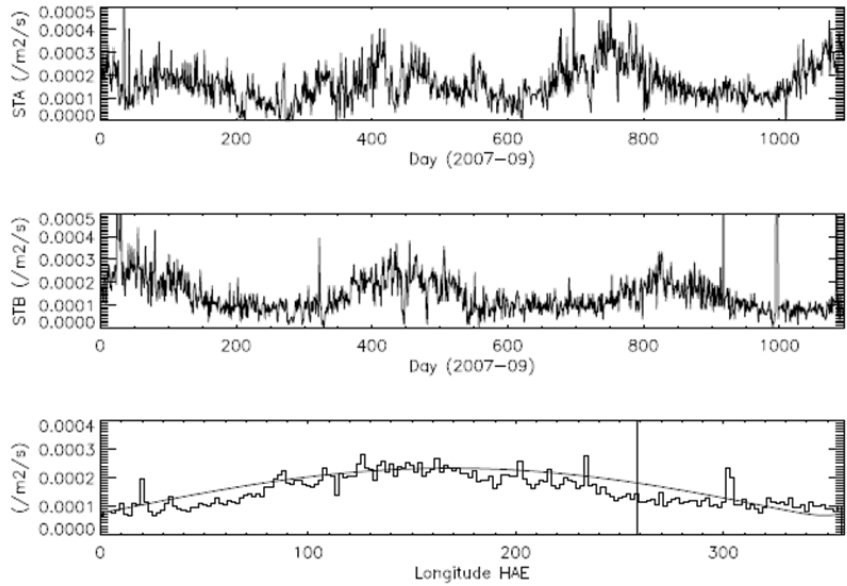

Figure 6: The top panels shows the flux of 0.1-0.3 μm dust grains as observed by the STEREO-A and STEREO-B spacecraft respectively, over the course of several years, showing an average flux of about $10^{-4}$ grains/m$^2$/s. Both spacecraft see a constant term, which is the uniform interplanetary component of the micron-sized dust, along with a modulated component with a one year period. The modulated component is a steady stream of interstellar dust flowing into the solar system in the direction of the Sun's motion through the Galaxy. The lower panel shows the dust rates observed by the two spacecraft as a function of longitude in the Heliocentric Aries ecliptic coordinate system, illustrating how the changing velocity vector of the spacecraft orbiting the Sun combines with the velocity of the solar system in interstellar space to modulate the rate of interstellar dust impacts.

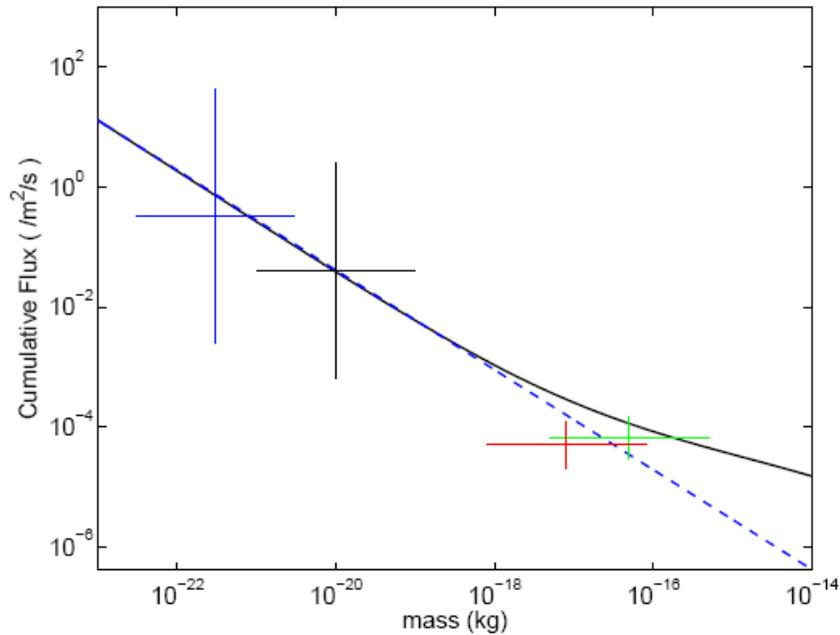

Figure 7: Comparison between the dust fluxes as a function of mass as determined though our analysis of the STEREO/WAVES radio observations (symbols) and an equilibrium model of interplanetary dust (solid and dashed lines, Grun et al, 1985).

# References

*Lunar Laser Ranging Constraints on the Lunar Core*

Battat, J. B. R., Murphy, T. W., Jr., Adelberger, E. G., Gillespie, B., Hoyle, C. D., McMillan, R. J., Michelsen, E. L., Nordtvedt, K., Orin, A. E., Stubbs, C. W., Swanson, H. E. 2009, "The Apache Point Observatory Lunar Laser-ranging Operation (APOLLO): Two Years of Millimeter-Precision Measurements of the Earth-Moon Range," Publications of the Astronomical Society of the Pacific, Vol. 121, No. 875, p. 29-40

Bender, P. L., Currie, D. G., Dicke, R. H., Eckhardt, D. H., Faller, J. E., Kaula, W. M., Mulholland, J. D., Plotkin, H. H., Poultney, S. K., Silverberg, E. C., Wilkinson, D. T., Williams, J. G., Alley, C. O. 1973, "The Lunar Laser Ranging Experiment," Science, Volume 182, Issue 4109, pp. 229-238

Chapront, J., Chapront-Touzé, M., Francou, G.A New determination of lunar orbital parameters, precession constant and tidal acceleration from LLR measurements, Astronomy and Astrophysics, v. 387, p.700-709

Currie, D. G., Dell'Agnello, S. & Delle Monache, G. O. 2011a, "A Lunar Laser Ranging Retroreflector Array for the 21st Century," Acta Astronautica, v. 68, iss. 7-8, p. 667-680

Currie, D.G., Dell'Agnello, S. & Delle Monache G.O. 2011b, "Lunar Laser Ranging Retroreflector for the 21st Century," 17th International Workshop on Laser Ranging, Proceedings of the conference held 16-20 May, 2011 in Bad Kotzing, Germany. To be published online at http://cddis.gsfc.nasa.gov/lw17

Currie, D. G. 2011a, "A Lunar Laser Ranging Retro-Reflector Array For The 21$^{st}$ Century," Google Lunar X Prize Science Workshop, NASA Lunar Science Institute, 13 July 2011

Currie, D. G. 2011b, "Deployment Aspects of the Lunar Laser Ranging Retroreflector for the 21st Century," NASA Lunar Science Institute/Lunar Science Workshop Forum, 19-21 July 2011

Currie, D. G. 2011c, "Anchored Lunar Deployment of the Lunar Laser Ranging Retroreflector for the 21st Century" NASA Lunar Science Institute/Lunar Science Workshop Forum, 20 July 2011

Currie, D. G., Horanyi, H., Murphy, T., Drake, K., Collette, A. & Shuand, A. 2011, "LIFETIME ISSUES FOR LUNAR RETROREFLECTORS: CCLDAS Dust Accelerator," NASA Lunar Science Institute/Lunar Science Workshop Forum, 19-21 July 2011

Dell'Agnello, S., Delle Monache, G. O., Currie, D. G., Vittori, R., Cantone, C., Garattini, M., Boni, A., Martini, M., Lops, C., Intaglietta, N., Tauraso, R., Arnold, D. A., Pearlman, M. R.,

Bianco G., et al, 2011, "ETRUSCO-2: an ASI-INFN Project of Development and SCF-Test of GNSS Retroreflector Arrays (GRA) for Galileo and the GPS-3," 17th International Workshop on Laser Ranging, Proceedings of the conference held 16-20 May, 2011 in Bad Kotzing, Germany. To be published online at http://cddis.gsfc.nasa.gov/lw17

Dell'Agnello, S., Delle Monache, G. O., Currie, D. G., Vittori, R.; Cantone, C., Garattini, M., Boni, A., Martini, M., Lops, C., Intaglietta, N., Tauraso, R., Arnold, D. A., Pearlman, M. R., Bianco, G., Zerbini, S., Maiello, M., Berardi, S., Porcelli, L., Alley, C. O., McGarry, J. F., Sciarretta, C., Luceri, V. & Zagwodzki, T. W. 2011, "Creation of the new industry-standard space test of laser retroreflectors for the GNSS and LAGEOS," Advances in Space Research, Volume 47, Issue 5, p. 822-842.

Murphy, Thomas W., Adelberger, E. G., Battat, J. B. R., Hoyle, C. D., McMillan, R. J., Michelsen, E. L., Stubbs, C. W., Swanson, H. E. 2009, "Lunar Laser Ranging: Science Achievements and Recent Advances," American Astronomical Society, IAU Symposium #261. Relativity in Fundamental Astronomy: Dynamics, Reference Frames, and Data Analysis 27 April - 1 May 2009 Virginia Beach, VA, USA, #8.02, Bulletin of the American Astronomical Society, Vol. 41, p.883

Pearlman, M. R., Degnan, J. J., Bosworth, J. M. 2002 "The International Laser Ranging Service", Advances in Space Research, Vol. 30, Issue 2, p.135-143

Rambaux, N., Williams, J. G. & Boggs, D. H. 2008, "A Dynamically Active Moon – Lunar Free Librations and Excitation Mechanisms," Lunar and Planetary Science Conference XXXIX, March 10-14, 2008.

Rambaux, N., & Williams, J. G. 2009, "The Moon's physical librations and determination of its free modes," Celestial Mechanics and Dynamical Astronomy, 2009.

Ratcliff, J. T., Williams, J. G. & Turyshev, S. G. 2008, "Lunar Science from Laser Ranging – Present and Future," Lunar and Planetary Science Conference XXXIX, March 10-14, 2008.

Williams, J. G., Boggs, D. H., Noda, H. 2010, "Exploring the Lunar Interior with Tides, Gravity and Orientation," American Astronomical Society, DPS meeting #42, #21.10; Bulletin of the American Astronomical Society, Vol. 42, p.987

Williams, J. G., & Boggs, D. H. 2009, "Lunar Core and Mantle. What Does LLR See?," Proceedings of 16th International Workshop on Laser Ranging, SLR – the Next Generation, October 2008, Poznan, Poland, ed. Stanislaw Schillak, 101-120, 2009, http://www.astro.amu.edu.pl/ILRS_Workshop_2008/index.php.

Williams, J. G., Turyshev S. G., & Folkner, W. M. 2010, "Lunar Geophysics and Lunar Laser Ranging, Ground-Based Geophysics on the Moon," Tempe, AZ, January 21-22, 2010

Zacny, K. & Currie, D. G. 2011a "Development and Testing of Gas Assisted Drill for the Emplacement of the Corner Cube Reflector System on the Moon," NASA Lunar Science Institute/Lunar Science Workshop Forum, 19-21 July 2011

*Riometry Measurements of the Lunar Ionosphere:*

Bauer, S. J. 1996, "Limits to a Lunar Ionosphere," Sitzungsberichte und Anzeiger, Abt. 2, 133, 17

Cane, H. 1979, "Spectra of the non-thermal radio radiation from the galactic polar regions," Mon. Not. R. Astron. Soc., 189, 465

Challis, J. 1863, "On the Indications by Phenomena of Atmospheres to the Sun, Moon and Planets," Mon. Not. R. Astron. Soc., 23, 231

Edwards, W. F., & Borst, L. B. 1958, "Possible Sources of a Lunar Atmosphere," Science, 127, 325

Elsmore, B. 1957, "Radio observations of the lunar atmosphere," Philos. Magazine, 2, 1040

Herring, J. R., & Licht, A. L. 1959, "Effect of the Solar Wind on the Lunar Atmosphere," Science, 130, 266

Hodges, R. R., Hoffman, J. H., & Johnson, F. S. 1974, "The Lunar Atmosphere," Icarus, 21, 415

Imamura, T., Iwata, T., Yamamoto, Z., et al. 2008, "Studying the Lunar Ionosphere with SELENE Radio Science Experiment," American Geophysical Union, #P51D-04

Johnson, F. S. 1971, "Lunar Atmosphere," Rev. Geophys. Space Phys., 9, 813

Milford, S. N., & Pomilla, F. R. 1967, "A Diffusion Model for the Propagation of Gases in the Lunar Atmosphere," J. Geophys. Res., 72, 4533

Pluchino, S., Schillirò, F., Salerno, E., Pupillo, G., Maccaferri, G., & Cassaro, P. 2008, "Radio Occultation Measurements of the Lunar Ionosphere," Memorie Soc. Astron. Italiana Suppl., 12, 53

Polisensky, E. 2007, "LFmap: A Low Frequency Sky Map Generating Program," LWA Memo #111; http://www.ece.vt.edu/swe/lwa/lwa0111.pdf

Reasoner, D. L., & O'Brien, B. J. 1972, J. Geophys. Res., 77, 1292

Rennilson, J. J., & Criswell, D. R.  1974, "Surveyor Observations of Lunar Horizon-Glow," Moon, 10, 121

Schroeter, J. J.  1792, "Observations on the Atmospheres of Venus and the Moon, Their Respective Densities, Perpendicular Heights, and the Twi-Light Occasioned by Them," Philos. Trans. R. Soc. London, 82, 309

Vyshlov, A. S.  1976, "Preliminary results of circumlunar plasma research by the Luna 22 spacecraft," Space Research, XVI, Proc. of Open Meetings of Workshop Groups of Physical Sciences (Akademie-Verlag) 945

Vyshlov, A. S., & Savich, N. A.  1978, "Observations of radio source occultations by the Moon and the nature of the plasma near the Moon," Cosmic Res., 16 (transl. Kosmicheskie Issledovaniya, 16), 551

Zook, H. A., & McCoy, J. E.  1991, "Large scale lunar horizon glow and a high altitude lunar dust exosphere," Geophys. Res. Lett., 18, 2117

*Nanodust Impacts:*

Grun et al. 1985, "Collisional balance of the meteoritic complex", Icarus 62, 244-272

Grun et al 1993, "Discovery of Jovian dust streams and interstellar grains by the Ulysses spacecraft", Nature 362, 428 - 430 (01 April 1993); doi:10.1038/362428a0

Meyer-Vernet, N., 2001. Detecting dust with electric sensors in planetary rings, comets and interplanetary space. ESA SP-476 (Edited by R. A. Harris), 635640.

Meyer-Vernet, N.,  et al. 2009, "Dust detection by the wave instrument on stereo: nanoparticles picked up by the solar wind?", Solar Physics 256, 463–474.

Zaslavsky et al 2011, "Interplanetary dust detection by radio antennas: mass calibration and fuxes measured by STEREO/WAVES", Planetary and Space Science, submitted.